\documentstyle[12pt]{article}
\begin{document}
\parskip=4mm

\newtheorem{prop}{Proposition}
\newtheorem{lem}{Lemma}
\renewcommand{\Re}{\mbox{\rm Re }}
\renewcommand{\Im}{\mbox{\rm Im }}
\newcommand{\C}{{\mathchoice{\setbox0=\hbox{$\displaystyle\sf C$}\hbox{\hbox
to0pt{\kern0.4\wd0\vrule height0.9\ht0\hss}\box0}}
{\setbox0=\hbox{$\textstyle\sf C$}\hbox{\hbox
to0pt{\kern0.4\wd0\vrule height0.9\ht0\hss}\box0}}
{\setbox0=\hbox{$\scriptstyle\sf C$}\hbox{\hbox
to0pt{\kern0.4\wd0\vrule height0.9\ht0\hss}\box0}}
{\setbox0=\hbox{$\scriptscriptstyle\sf C$}\hbox{\hbox
to0pt{\kern0.4\wd0\vrule height0.9\ht0\hss}\box0}}}}
\newcommand{\ddoteq}{\,\raisebox{-2mm}{$\stackrel{\displaystyle \doteq}
 {\mbox{\raisebox{1mm}{.}}}$}}

\begin{center}
{\large\bf Determination of the chiral pion-pion scattering 
parameters:\\ a proposal}

{\bf G\'erard Wanders}

{\small Institut de physique th\'eorique, Universit\'e de Lausanne,\\ 
CH-1015 Lausanne, Switzerland}
\end{center}

\abstract{An explicitly crossing-symmetric decomposition of the pion-pion 
scattering amplitudes into low- and high-energy components is established. The 
high-energy components are entirely determined by absorptive parts at high 
energies. With the exception of the two leading-order parameters, all the 
parameters appearing in the one- and two-loop chiral amplitudes are determined 
by the high-energy components of the exact amplitudes.}

\section{Introduction and statement of results}

Chiral perturbation theory of the meson sector is an effective field theory 
providing a successful description of low-energy strong interaction processes 
in terms of expansions in powers of external momenta and quark 
masses~\cite{Gas}. These expansions are derived from an effective Lagrangian 
which is itself a series of powers of the pion field, its derivatives and the 
quark mass matrix. It contains effective coupling constants whose number 
increases dramatically as one proceeds from the leading terms to higher order 
corrections~\cite{Fea}. The problem of determining the values of 
these coupling constants arises. As one is dealing with an effective 
low-energy theory, part of the coupling constants is meant to encode the 
low-energy manifestations of high-energy phenomena, therefore it must be 
possible to relate some of the coupling constants to the characteristics of 
such phenomena. In the present context, high energies are of course very 
modest, starting around 500~MeV. Order of magnitude 
estimates can be obtained by evaluating contributions of the high-energy 
states (resonances), by means of a Lagrangian 
describing these states and their coupling to the pion field~\cite{Eck}. This 
procedure amounts to saturating high-energy cross-sections by resonance 
contributions in a narrow width approximation. Other, potentially more 
accurate, methods are based on the use of dispersion relations which connect 
low- and high-energy processes~\cite{Kne1,Pen}.

In this paper I follow the dispersion relation path and consider a special 
process, pion-pion scattering. I address the problem of determining the 
parameters appearing in the chiral amplitudes of this process. These 
parameters are related in a known way to the coupling constants of the chiral 
Lagrangian. More precisely, I am asking two questions:
\begin{enumerate}
\item Is it possible to decompose a pion-pion amplitude into a high- and a 
low-energy component? The high-energy component should be determined by 
high-energy absorptive parts and it should be possible to obtain the 
low-energy component from the chiral absorptive part.
\item Is it possible to determine unambiguously the chiral parameters and, 
consequently, the chiral coupling constants, with the aid of the high-energy 
components?
\end{enumerate}

As crossing symmetry is a basic property of pion-pion scattering, I require 
the decomposition into low- and high-energy components to be explicitly 
crossing symmetric. If one works with dispersion relations, the main 
difficulty of question (1) comes from this last requirement. In fact, ordinary 
dispersion relations are not convenient tools and a technique developed thirty 
years ago turns out to be more appropriate~\cite{Wan}. It is based on the 
following considerations. The isospin $I$ $s$-channel amplitude $T^I$ is a 
function of the three Mandelstam variables $s$, $t$ and $u$ 
($I=0,1,2$)~\cite{Ros,Mah}. Crossing symmetry dictates the transformation of 
the $T^I$ under permutation of $s$, $t$ and $u$. One defines three amplitudes 
$G_i$, linearly related to the $T^I$, which are totally symmetric functions of 
$s,t,u$ ($i=0,1,2$). Conversely the total symmetry of the $G_i$ implies 
crossing symmetry for the $T^I$. When expressed in terms of appropriate new 
variables the $G_i$ obey dispersion relations which do not spoil their symmetry 
properties. The low- and high-energy components $L_i$ and $H_i$ of $G_i$ are 
obtained by splitting its dispersion integral into low- and high-energy parts. 
The $L_i$ and $H_i$ are totally symmetric and define a crossing symmetric 
decomposition of the $T^I$. Therefore the answer to question~(1) is affirmative.

A strategy for fixing the chiral parameters is to adjust them in such a way 
that the chiral amplitudes $T_\chi^I$ are good approximations of the true 
amplitudes $T^I$ at points where the chiral expansion has to be valid. I adopt 
and implement this strategy by requiring that truncated Taylor expansions of 
$T^I$ and $T_\chi^I$ coincide at a conveniently chosen point where both 
amplitudes are regular. Points in the Mandelstam triangle $s<4M_\pi^2$, 
$t<4M_\pi^2$, $u<4M_\pi^2$ are good candidates and I shall work with Taylor 
expansions around the symmetry point $s=t=u=4M_\pi^2/3$. The outcome will be a 
set of equations relating the parameters of the higher-order terms of the 
chiral amplitudes to high-energy pion-pion scattering, two leading-order 
parameters remaining unconstrained. Consequently the answer to question~(2) is 
also affirmative to a large extent: it confirms the special status of the 
leading-order coupling constants.

As a by-product of my investigations I obtain upper bounds for the Taylor 
coefficients of the high-energy components at the symmetry point. They seem to 
be compatible with good convergence properties of the chiral expansion.

A similar method for the determination of pion-pion parameters has been 
developed in~\cite{Kne1}. The main difference between this method and my 
proposal lies in a treatment of crossing symmetry which does not depend on the 
order of the chiral expansion.

The principal aim of this paper is to establish that the idea of determining 
chiral coupling constants from high-energy processes can be realized precisely 
and unambiguously in the special case of pion-pion scattering. My technique 
cannot be extended to other processes in a straightforward way. Here, I am 
mainly interested in questions of principle and the practical application of my 
constraints is another task. Due to the poor shape of our information on 
high-energy pion-pion scattering, it is doubtful that they really can improve 
the results already obtained~\cite{Kne1}.

The paper is organized as follows. Section~2 contains an outline of 
the derivation of dispersion relations for the totally symmetric amplitudes 
$G_i$. Section~3 is technical: the analyticity properties allowing Taylor 
expansions in two variables around the symmetry point are established. The 
results are stated in two propositions. Constraints for the chiral coupling 
constants are derived in Section~4. Explicit equations up to the sixth order of 
the chiral expansion are written down. Technical details are presented in two 
Appendices.

\section{Dispersion relations for totally symmetric \hfill\break amplitudes}

High energy components of the pion-pion amplitudes $T^I$ will be defined with 
the help of three totally symmetric functions $G_i(s,t,u)$:
\begin{eqnarray}
G_0(s,t,u) &=& {1\over 3}\left(T^0(s,t,u)+2T^2(s,t,u)\right),\nonumber\\
G_1(s,t,u) &=& {T^1(s,t,u)\over t-u}+{T^1(t,u,s)\over u-s}+
{T^1(u,s,t)\over s-t},\nonumber\\
G_2(s,t,u) &=& {1\over s-t}\left({T^1(s,t,u)\over t-u}-{T^1(t,s,u)\over s-
u}\right)\label{2eq1}\\
&&+\,{1\over t-u}\left({T^1(t,u,s)\over u-s}-{T^1(u,t,s)\over t-s}\right)
+\,{1\over u-s}\left({T^1(u,s,t)\over s-t}-{T^1(s,u,t)\over u-t}\right).\nonumber
\end{eqnarray}

The Mandelstam variables will be expressed in units of $M_\pi^2$, $M_\pi =$ 
pion mass ($s+t+u=4$). No poles are produced by the 
denominators because $T^1(s,t,u)$ is antisymmetric in $t$ and $u$. The 
functions $G_0$, $G_1$ and $G_2$ have been introduced by Roskies~\cite{Ros}: 
$G_0$ is simply the $\pi^0-\pi^0$ amplitude. Crossing symmetry is encoded in 
the total symmetry of the $G_i$. The individual amplitudes $T^I$ are 
reconstructed in the following way.
\begin{eqnarray}
T^0(s,t,u)&=&{5\over 3}G_0(s,t,u)+{2\over 9}(3s-4) G_1(s,t,u)-{2\over 
27}[3s^2+6tu-16]G_2(s,t,u), \nonumber\\
T^1(s,t,u)&=&(t-u)\left[{1\over 3}G_1(s,t,u)+{1\over 9}(3s-
4)G_2(s,t,u)\right],\label{2eq2}\\
T^2(s,t,u)&=&{2\over 3}G_0(s,t,u)-{1\over 9}(3s-4) G_1(s,t,u)+{1\over 
27}[3s^2+6tu-16]G_2(s,t,u).\nonumber
\end{eqnarray}

The symmetry of $G_i$ implies that it can be expressed as a function of two 
independent variables which are totally symmetric and homogeneous in $s$, $t$ 
and $u$, for instance the variables $x$ and $y$ defined by
\begin{equation}\label{2eq3}
x=-{1\over 16}(st+tu+us),\qquad y={1\over 64}stu.
\end{equation}
No singularities are induced by the change of variables $(s,t,u)\to (x,y)$: 
each singularity of $G_i$ as a function of $x$ and $y$ is the image of a 
singularity in the $(s,t,u)$-space. Analyticity properties of $G_i(x,y)$ have 
been established~\cite{Wan,Mah} by looking at its restrictions to 
complex straight lines
\begin{equation}\label{2eq4}
y=a(x-x_0)+y_0,\qquad a,x_0,y_0\in\C.
\end{equation}

As a function of $x$, at a fixed value of the slope $a$ and for a given point 
$(x_0,y_0)$, the restriction
\begin{equation}\label{2eq5}
F_i(x;a,x_0,y_0)\,\ddoteq \,G_i(x,a(x-x_0)+y_0)
\end{equation}
has simple analyticity properties. As long as $a$ and $y_0$ belong to an 
$x_0$-dependent neighborhood $V(x_0)$ of the origin, $F_i$ is regular in the 
$x$-plane with a cut $C(a,x_0,y_0)$. This cut is the image in the $x$-plane of 
the physical cut $\{s,t,u\,|\,4\leq s<\infty\}$. The Froissart bound for the 
asymptotic behavior of the pion-pion amplitudes implies a once subtracted 
dispersion relation for $F_0$ and $F_1$ and an unsubtracted relation for 
$F_2$:
\begin{eqnarray}
\lefteqn{F_i(x; a,x_0,y_0)-(1-\delta_{i2})F_i(x_1,a,x_0,y_0)\qquad}\nonumber\\[-2mm]
&&\label{2eq6}\\[-2mm]
\hspace{1cm}&=&{1\over \pi}\int_{C(a,x_0,y_0)}{\rm d}\xi\left[{1\over \xi-x}-
(1-\delta_{i2}){1\over \xi-x_1}\right]{\rm Disc}\,F_i(\xi,a,x_0,y_0),\nonumber
\end{eqnarray}
where Disc~$F_i$ is the discontinuity of $F_i$ across the cut $C$. The 
relation (\ref{2eq6}) holds whenever $(a,y_0)\in V(x_0)$; it is an exact 
consequence of the general principles of quantum field theory.

My high-energy components of the pion-pion amplitudes will be obtained from a 
decomposition of the right-hand side integral in (\ref{2eq6}) into low- and 
high-energy parts. To this end, it is convenient to parameterize the cut $C$ by 
means of the energy squared $s$ and rewrite the right-hand side of (\ref{2eq6}) 
as an integral over $s$. The variable $\xi$ in (\ref{2eq6}) becomes a function 
of $s$:
\begin{equation}\label{2eq7}
\xi(s;a,x_0,y_0)={1\over 16(s+4a)}\left[s^2(s-4)+64(ax_0-y_0)\right].
\end{equation}

The discontinuity of $F_i$ is related to the absorptive parts $A^I(s,t)$ of 
the pion-pion amplitudes evaluated at a (complex) value $\tau(s;a,x_0,y_0)$ of 
the squared momentum transfer $t$:
\begin{equation}\label{2eq8}
\tau(s;a,x_0,y_0)=-{1\over 2}\left\{(s-4)-\left[(s-4)^2-{16\over 
s+4a}\left(as(s-4)-16(ax_0-y_0)\right)\right]^{1\over 2}\right\}.
\end{equation}

With these prerequisites the change of variable $\xi\to s$ transforms 
(\ref{2eq6}) into
\begin{equation}\label{2eq9}
G_i(x,y)=(1-\delta_{i2})G_i(x_1,y_1)+{1\over 16\pi}\int_4^\infty{\rm 
d}s{1\over s+4a}\left[{1\over \xi-x}-(1-\delta_{i2}){1\over \xi-
x_1}\right]B_i(s,\tau).
\end{equation}
The relation (\ref{2eq6}) has been written in terms of $G_i$, the points $(x,y)$ and 
$(x_1,y_1)$ belonging to the straight line $(2,4)$. In the integral, $\xi$ and 
$\tau$ denote the functions defined in (\ref{2eq7}) and (\ref{2eq8}). The 
function $B_i$ is proportional to Disc~$F_i$:
\begin{equation}\label{2eq10}
B_i(s,\tau)=(s-\tau)(2s-4+\tau){\rm Disc}~F_i(\xi;a,x_0,y_0).
\end{equation}
It is obtained from the absorptive parts $A^I$:
\begin{eqnarray}
B_0(s,t)&=&{1\over 3}(s-t)(2s-4+t)\left(A^0(s,t)+A^2(s,t)\right), \nonumber\\
B_1(s,t)&=&{1\over 6}(3s-4)\left(2A^0(s,t)-5A^2(s,t)\right) \nonumber\\
&&\qquad +\left[{(s-t)(2s-4+t)\over (2t-4+s)}-{1\over 2}(2t-
4+s)\right]A^1(s,t),\label{2eq11}\\
B_2(s,t)&=&-{1\over 2}\left(2A^0(s,t)-5A^2(s,t)\right)+{3\over 2}{3s-4\over 
2t-4+s}A^1(s,t).\nonumber
\end{eqnarray}
The construction of the domain $V(x_0)$ specifying the validity of 
relation~(\ref{2eq9}) can now be explained. It is based on the known exact 
properties of absorptive parts~\cite{Mar}. At fixed $s$ ($4\leq s<\infty$) 
$A^I(s,t)$ is an analytic function of $t$ which is regular in an ellipse 
$E(s)$ with foci at $t=0$ and $t=-(s-4)$ and right extremity $r(s)$ given by
\begin{equation}\label{2eq12}
r(s)=\left\{\begin{array}{lll}
\displaystyle{16s\over s-4} &\qquad \mbox{for}\quad & 4<s<16,\\[4mm]
\displaystyle{256\over s} &\qquad \mbox{for}\quad & 16\leq s\leq 32,\\[4mm]
\displaystyle{4s\over s-16} &\qquad \mbox{for}\quad & 32\leq s<\infty.
\end{array}\right.
\end{equation}

The integrand in (\ref{2eq9}) is defined if $\tau(s;a,x_0,y_0)$ always stays 
within the ellipse $E(s)$. This is precisely the condition defining the domain 
$V(x_0)$ which I shall use:
\begin{equation}\label{2eq13}
V(x_0)=\left\{a,y_0\,\mid\,\tau(s;a,x_0,y_0)\in E(s),\quad 4\leq s<\infty
\right\}.
\end{equation}
Figure~1 illustrates the limitations resulting from the condition
\begin{equation}\label{2eq14}
(a,y_0)\in V(x_0)
\end{equation}
for real values of the parameters $a$, $x_0$ and $y_0$. Figure~2 displays the 
permitted values of the slope $a$ when $x_0=-50$ and $y_0=1/27$. If (\ref{2eq14}) is 
fulfilled, the relation (\ref{2eq9}) not only holds true but the absorptive 
parts appearing in $B_i$ are given by their convergent partial wave 
expansions. In this sense the integral in (\ref{2eq9}) only involves physical 
quantities.

\section{Defining high-energy components}

From now on, I keep $x_0$ and $y_0$ fixed, and confine myself to the family of 
straight lines (\ref{2eq4}) passing through the point 
$(x_0,y_0)$. Furthermore, the subtraction point $(x_1,y_1)$ is identified with 
$(x_0,y_0)$ and the dispersion relation (\ref{2eq9}) is used as a representation 
of the function $G_i(x,y)$ in a domain of the $(x,y)$-space determined by 
condition (\ref{2eq14}) with $a=(y-y_0)/(x-x_0)$. I choose $(x_0,y_0)$ 
in such a way that this representation holds in a neighborhood of the 
point $x=x_s=-1/3$, $y=y_s=1/27$, which is the image of the symmetry point 
$s=t=u=4/3$. The Taylor expansion of $G_i$ around this point can then be 
obtained from the representation and the parameters of the chiral pion-pion 
amplitudes are constrained by equating Taylor coefficients of the chiral $G_i$ 
with the coefficients derived from (\ref{2eq9}). This explicitly crossing 
symmetric procedure will be explained in detail in Section~4.

The main aim of the present Section is the extraction of high-energy 
components from equation~(\ref{2eq9}), but it is first necessary to ensure that 
this equation really provides a representation of $G_i$ in a neighborhood of 
the symmetry point.

\begin{prop} \label{3prop1}If $x_0$ is real, $-72<x_0<3x_s/2$, $y_0=y_s$, the 
function $G_i(x,y)$ given by equation (\ref{2eq9}) is regular for $(x,y)\in M$ 
where $M$ is the cartesian product $D_x\times D_y$ of two disks $D_x$ and 
$D_y$ in the $x$- and $y$-plane respectively, centered on $x_s$ and $y_s$.
\end{prop}

The integral $J(x,a)$ in (\ref{2eq9}) is primarily a function of $x$ and $a$. 
Since $x_0$ and $y_0$ are fixed, condition (\ref{2eq14}) defines a domain $W$ 
for $a$. The integral $J(x,a)$ is defined and regular if $a$ belongs to 
$W$ and $x$ is in $\C\backslash C(a)$, $C(a)$ being an abbreviation for 
$C(a,x_0,y_0)$ (the point $(x_0,y_0)$ being fixed, explicit reference to the 
$x_0$- and $y_0$-dependence will also be dropped in $\xi$ and $\tau$). 
Information on the location of the cut $C(a)$ is needed in order to proceed. 
If the slope $a$ is real, inspection of Fig.~1 shows that
\begin{equation}\label{3eq1}
\xi(s,a)\geq \xi(4,a_0)
\end{equation}
if $s\geq 4$ and $-1<a\leq a_0$. This means that for such values of $a$, the 
cut $C(a)$, which is on the real $x$-axis, is entirely on the right of the 
point $x=\xi(4,a_0)=(a_0x_0-y_0)/(1+a_0)$.

An inequality similar to (\ref{3eq1}) holds for complex slopes and in a more 
general context.

\begin{lem}\label{3lem1} If $|a|<a_0$, $a_0<\Lambda^2/4$, $\Lambda^2\geq 4$, $y_0>0$ and 
$x_0+y_0<0$, the inequality
\begin{equation}\label{3eq2}
\Re \xi(s,a)>\xi(\Lambda^2,a_0)
\end{equation}
holds for $\Lambda^2\leq s<\infty$.
\end{lem}

This lemma follows from a straightforward computation.

Setting $\Lambda^2=4$ in (\ref{3eq2}) one sees that the whole complex cut 
$C(a)$ is on the right of the line $\Re x=\xi(4,a_0)$.

\begin{lem}\label{3lem2} The integral $J(x,a)$ is defined and regular for
$\Re x<\xi(4,a_0)$ and $|a|<a_0$ if $x_0<3x_s/2$ and $a_0<1$. The circle 
$|a|=a_0$ has to be inside the domain $W$.
\end{lem}

To prove this lemma I choose $a_0$ in such a way that $\xi(4,a_0)=x_s/2=-1/6$. With $y_0=y_s$ 
this gives
\begin{equation}\label{3eq3}
a_0=-{x_s+2y_s\over x_s-2x_0}.
\end{equation}
As, by assumption, $x_0<3x_s/2$, the condition $a_0<1$ is satisfied. One 
verifies that the circle $|a|=a_0$ is contained in $W$ if 
$-72<x_0<3x_s/2$. Consequently, Lemma~\ref{3lem2} shows that $J(x,a)$ 
is regular in the product $D_x\times D_a$ of two disks, $D_x$ with center 
$x=x_s$ and radius $\rho_x=-x_s/2$, and $D_a$ with center $a=0$ and radius 
$a_0$ given in (\ref{3eq3}). This result implies Proposition~\ref{3prop1}. 
Indeed, the representation (\ref{2eq9}) can be rewritten as
\begin{equation}\label{3eq4}
G_i(x,y)=G_i(x_0,y_0)+J\left(x,{y-y_0\over x-x_0}\right).
\end{equation}
If $x\in D_x$, $y\in D_y$, the radius of $D_y$ being $\rho_y=a_0((3x_s/2)-
x_0)$, the slope $a=(y-y_s)/(x-x_0)$ verifies the inequality
\begin{equation}\label{3eq5}
|a|={|y-y_s|\over |x-x_0|}<a_0,
\end{equation}
because $|y-y_s|<\rho_y$ and $|x-x_0|>(3x_s/2)-x_0$ when $|x-x_s|<\rho_x$. 
Therefore $a\in D_a$ and the right-hand side of (\ref{3eq4}) is defined and 
regular.\hfill{$\Box$}

The proof of Proposition~\ref{3prop1} contains arbitrary choices leading to 
special, non-optimal, values of the radii $\rho_x$ and $\rho_y$. This does 
not matter because the role of Proposition~\ref{3prop1} is simply to ensure 
analyticity in a product of two disks which, in turn, guarantees the 
convergence of the Taylor expansion of $G_i$ in both variables $x$ and $y$.

The validity of (\ref{3eq4}) in a neighborhood of the symmetry point being 
established, a decomposition of $G_i$ into a 
low- and high-energy component valid in that neighborhood can be defined 
simply by splitting $J(x,a)$ into an integral from 4 to $\Lambda^2$ and an 
integral from $\Lambda^2$ to infinity. The high-energy component$H_i$ of 
$G_i$ is defined by
\begin{equation}\label{3eq6}
H_i(x,y)={1\over 16\pi}\int_{\Lambda^2}^\infty{\rm d}s{1\over 
(s+4a)}\left[{1\over \xi-x}-(1-\delta_{i2}){1\over \xi-x_0}\right]B_i(s,\tau)
\end{equation}
where $a=(y-y_0)/(x-x_0)$. The following proposition, an analogue of 
Proposition~\ref{3prop1}, holds for $H_i$.

\begin{prop}\label{3prop2} If 
\begin{equation}\label{3eq7}
-72<x_0<-{1\over 64}\Lambda^4(\Lambda^2-4)+\left({1\over 
4}\Lambda^2+1\right)x_s+y_s
\end{equation}
the function $H_i$ defined in (\ref{3eq6}) is regular in the union 
$M_H$ of a family of cartesian products of two disks:
\begin{equation}\label{3eq8}
M_H=\bigcup_{\rho_x}\left[D_x(\rho_x)\times
D_y(\rho_y)\right].
\end{equation}
The disks $D_x$ and $D_y$ are centered at $x=x_s$ and $y=y_s$ and 
their radii $\rho_x$ and $\rho_y$ are related by
\begin{equation}\label{3eq9}
\rho_y={1\over 64}{x_s-\rho_x-x_0\over x_s+\rho_x-
x_0}\left[\Lambda^4(\Lambda^2-4)-16\Lambda^2(x_s+\rho_x)-64 y_s\right],
\end{equation}
with
\begin{equation}\label{3eq10}
0< \rho_x<{\Lambda^2\over 16}(\Lambda^2-4)-x_s-{4 y_s\over 
\Lambda^2}.
\end{equation}
\end{prop}

The proof of Proposition~\ref{3prop2} is a paraphrase of the proof of 
Proposition~\ref{3prop1}. If $J_H(x,a)$ denotes the integral in 
(\ref{3eq6}), this function is defined if $a$ belongs to a domain $W_H$ 
obtained from (\ref{2eq13}) by restricting $s$ to values larger than 
$\Lambda^2$. As a function of $x$ $J_H(x,a)$ has a cut $C_H(a)$:
\begin{equation}\label{3eq11}
C_H(a)=\left\{ x\,\mid\,x=\xi(s,a),\quad s\geq\Lambda^2\right\}.
\end{equation}
Lemma~\ref{3lem1} now indicates that the cut $C_H(a)$ is entirely on the 
right of the line $\Re x=\xi(\Lambda^2,a_0)$ if $|a|<a_0<(\Lambda^2/4)$, 
$\xi(\Lambda^2,a_0)$ being given by (\ref{2eq7}). This quantity is the 
abscissa of the intersection of the line $y=a_0(x-x_0)+y_s$ with the image
\begin{equation}\label{3eq12}
64 y=\Lambda^2\left(\Lambda^2(\Lambda^2-4)-16x\right)
\end{equation}
of the line $s=\Lambda^2$ in the real $(x,y)$-plane.

By analogy with Lemma~\ref{3lem2}, it now appears that the integral $J_H$ 
is defined and regular for $\Re x<\xi(\Lambda^2,a_0)$ and $|a|<a_0$, provided 
that the circle $|a|=a_0$ is within $W_H$. If $x_0$ verifies (\ref{3eq7}) this 
last condition is satisfied for $a_0<1$.

The regularity of $H_i(x,a)$ for $x\in D_x$ is ensured if the radius 
$\rho_x$ is such that
\begin{equation}\label{3eq13}
\rho_x<\xi(\Lambda^2,a_0)-x_s.
\end{equation}

At a given $\rho_x$ this fixes the maximal slope $a_0$:
\begin{equation}\label{3eq14}
a_0={\Lambda^4(\Lambda^2-4)-16\Lambda^2(x_s+\rho_s)-64 y_s\over
64(x_s+\rho_x-x_0)}.
\end{equation}
As $a_0$ must be positive, $\rho_x$ has the upper bound (\ref{3eq10}).
Furthermore, $a_0$ has to be smaller than 1 in order to secure regularity 
with respect to $a$ in $D_a$, the disk $|a|<a_0$, $a_0$ given by 
(\ref{3eq14}). If $\rho_x$ is allowed to vanish, this imposes the upper limit 
in (\ref{3eq7}). As in the last step of the proof of Proposition~\ref{3prop1}, 
the regularity of $J_H(x,a)$ for $(x,a)\in D_x\times D_a$ now 
implies the regularity of $H_i(x,y)$ in $D_x(\rho_x)\times 
D_y(\rho_y)$, the radius of $D_y$ being 
$\rho_y=a_0(x_s-\rho_x-x_0)$. The expression (\ref{3eq14}) for 
$a_0$ leads to the relation (\ref{3eq9}) between $\rho_x$ and 
$\rho_y$.\hfill{$\Box$}

As the radii of convergence of the Taylor expansion of $H_i$ will matter, 
Proposition~\ref{3prop2} goes into greater detail than 
Proposition~\ref{3prop1} although it is not aimed at being optimal.

Whereas (\ref{3eq6}) defines crossing-symmetric high-energy components of the 
$T^I$ via (\ref{2eq2}), a drawback of these components is that they depend on 
the choice of the subtraction point $(x_0,y_0)$. This comes from the explicit 
appearance of $x_0$ in the integral of (\ref{3eq6}) (if $i\neq 2$) and the 
$(x_0,y_0)$-dependence of $\xi$ and $\tau$ (cf.~(\ref{2eq7}) and 
(\ref{2eq8})). In fact, after identification of $(x_0,y_0)$ and $(x_1,y_1)$ in 
(\ref{2eq9}), the right-hand side has to be independent of $(x_0,y_0)$ and 
this leads to constraints on the absorptive parts already noticed 
in~\cite{Mah}. If the integral is split into a low- and a high-energy part, 
there is a coupling between low- and high-energy absorptive parts, which I 
shall not discuss.

\section{High-energy constraints on the one- and two-loop chiral pion-pion 
parameters}

The outcome of the previous Sections is a representation of the symmetric 
amplitudes $G_i(x,y)$ in a neighborhood of the symmetry point. It provides a 
decomposition into low- and high-energy contributions,
\begin{equation}\label{4eq1}
G_i(x,y)=L_i(x,y)+H_i(x,y),
\end{equation}
where
\begin{equation}\label{4eq2}
L_i(x,y)=(1-\delta_{i2})G_i(x_0,y_0)+{1\over \pi}\int_4^{\Lambda^2}{\rm 
d}s{1\over s+4a}\left[{1\over \xi-x}-(1-\delta_{i2}){1\over \xi-x_0}\right] 
B_i(s,\tau)
\end{equation}
is the low-energy component and $H_i$ the high-energy component defined in 
(\ref{3eq6}). Proposition~\ref{3prop1} applies to $L_i$: this function is 
known to be regular in the domain $M$ of Proposition~\ref{3prop1}. The 
high-energy component $H_i$ is certainly regular in the larger domain $M_H$ 
defined in (\ref{3eq8}). These analyticity properties imply that the Taylor 
expansion of $G_i(x,y)$ around $(x_s,y_s)$ in the two complex variables $x$ 
and $y$ can be extracted from the representations (\ref{3eq6}) and 
(\ref{4eq2}). It converges in a domain containing $M$ whereas the expansion of 
the high-energy component converges in a larger domain containing $M_H$.

In order to derive well defined constraints on the parameters appearing in the 
chiral amplitudes $T_\chi^I$ from (\ref{4eq1}), I make two assumptions:
\begin{enumerate}
\item[(i)] The symmetric amplitudes $G_i^\chi$ obtained from the $2n$-th order 
chiral amplitudes approximate the true symmetric amplitudes $G_i$ in a 
neighborhood of the symmetry point up to higher order corrections.
\item[(ii)] The discontinuities Disc~$G_i^\chi$ of the $2n$-th chiral 
symmetric amplitudes approximate Disc~$G_i$ in a bounded interval above 
threshold up to higher order corrections.
\end{enumerate}
This means that the representation (\ref{4eq1}) can be rewritten in the 
following way if $\Lambda^2$ is conveniently chosen and if $(x,y)$ is close to 
$(x_s,y_s)$:
\begin{equation}\label{4eq3}
G_i^\chi(x,y)=L_i^\chi(x,y)+H_i(x,y)+\mbox{ higher order terms}.
\end{equation}
The low-energy component $L_i^\chi$ is obtained from (\ref{4eq2}) where $B_i$ 
is replaced by $B_i^\chi$.

The precise value of $\Lambda^2$ plays no role in what follows. A special 
value I have in mind is $\Lambda^2=16$ corresponding to an energy of 560~MeV.

Each chiral amplitude $T_\chi^I$ is a sum of a polynomial in $s$, $t$ and $u$ 
and non-polynomial terms exhibiting the cuts necessarily present in any 
scattering amplitude. According to (\ref{2eq1}) the symmetric amplitudes 
$G_i^\chi$ have the same structure. The polynomial part of $G_i^\chi$ is 
$O(p^{2n_i})$ where $n_i$ is determined by $n$ and depends on $i$. Although the 
$G_i^\chi$ do not have the same asymptotic behavior as the $G_i$, they share 
the regularity properties we have established. The coefficients appearing in 
the polynomials and in the non-polynomial terms are determined by the chiral 
coupling constants.

By construction $G_i^\chi$ and $L_i^\chi$ have the same discontinuity across 
the cut $C(a)$ as long as $4\leq s\leq \Lambda^2$. This implies that 
the difference $\left(G_i^\chi-L_i^\chi\right)$ can be written as
\begin{equation}\label{4eq4}
G_i^\chi(x,y)-L_i^\chi(x,y)=-(1-\delta_{i2})G_i(x_0,y_0)+P_i(x,y)+H_i^\chi(x,y)
\end{equation}
where $P_i$ is a low-energy component and $H_i^\chi$ is the high-energy 
component of $G_i^\chi$. As $P_i$ has no discontinuity across the low-energy 
part of the cut $C(a)$, it is regular in a domain which is larger than $M$. Up 
to the sixth order of the chiral expansion $P_i$ is in fact a polynomial of 
degree $2n_i$. The following discussion applies to that situation, i.e. I 
assume that $n\leq 3$ from now on. I show in Appendix~A how $P_i$ and 
$H_i^\chi$ are constructed.

Combining (\ref{4eq4}) and (\ref{4eq3}) gives
\begin{equation}\label{4eq5}
P_i(x,y)-(1-\delta_{i2})G_i(x_0,y_0)=H_i(x,y)-
H_i^\chi(x,y)+O\left(p^{2(n_i+1)}\right).
\end{equation}
This relation has to hold in a neighborhood of the symmetry point. It becomes 
a strict equality if $H_i$ and $H_i^\chi$ are replaced by their $2n_i$-order 
truncated Taylor expansions $Q_i$ and $Q_i^\chi$
\begin{equation}\label{4eq6}
P_i(x,y)+Q_i^\chi(x,y)=-(1-\delta_{i2})G_i(x_0,y_0)+Q_i(x,y).
\end{equation}
The left-hand side is entirely determined by the $2n$-th order chiral 
amplitudes whereas the right-hand side is fixed by the pion-pion absorptive 
parts above $\Lambda^2$ and the value of $G_i$ at the subtraction point if 
$i=0,1$. Equating the coefficients of the left- and right-hand side 
polynomials gives a series of constraints on the parameters of the chiral 
amplitudes. The $i=0$ and $i=1$ constraints coming from the constant terms in 
(\ref{4eq6}) have a special status because of the presence of $G_i(x_0,y_0)$, 
the unknown value of $G_i$ at the subtraction point. The remaining constraints 
relate the chiral parameters to high-energy pion-pion scattering.

The regularity of $H_i$ in the family of products $M_H$ implies upper bounds 
for the coefficients $C_{n,m}$ of its Taylor expansion
\begin{equation}\label{4eq7}
H_i(x,y)=\sum_{n,m}C_{n,m}^i(x-x_s)^n(y-y_s)^m.
\end{equation}

If $\rho_x$ and $\rho_y$ are such that $D_x(\rho_x)\times D_y(\rho_y)$ is 
inside $M_H$, $|H_i|$ is finite on the boundary of this product of two disks 
and 
\begin{equation}\label{4eq8}
\left|C_{n,m}^i\right|<{K_i\over (\rho_x)^n(\rho_y)^m}.
\end{equation}
One checks that $\rho_x=\rho_y=9$ fulfills the above requirements if $x_0=-50$ 
and $\Lambda^2=16$ (notice that these values are compatible with 
(\ref{3eq7})). This leads to the simple but severe bound 
$|C_{n,m}^i|<K_i/9^{(n+m)}$. A more refined bound is derived in Appendix~B. The 
same bounds hold for the Taylor coefficients of $H_i^\chi$. Inequality 
(\ref{4eq8}) is an important result. The exponential falloff of the 
$C_{n,m}^i$ when $n$ and/or $m$ increase indicates that a rapid decrease of 
the size of the high-order terms in the chiral expansion is conceivable.

Finally I examine the nature of the conditions that equation (\ref{4eq6}) 
imposes on the sixth-order, one- and two-loop chiral 
amplitudes~\cite{Kne2,Bij}. They are 
obtained in a standard way from a single function $A^\chi(s,t,u)$ which is the sum 
of a second-, fourth- and sixth-order term
\begin{equation}\label{4eq9}
A^\chi(s,t,u)=\lambda^2A_2(s,t,u)+\lambda^4A_4(s,t,u)+\lambda^6A_6(s,t,u)
\end{equation}
with $\lambda=M_\pi/F_\pi$. The polynomial parts of these terms have the form
\begin{equation}\label{4eq10}
\begin{array}{lll}
A_2^{\rm pol}(s,t,u)&=& a_{2,0}+a_{2,1}s\\
A_4^{\rm pol}(s,t,u)&=& a_{4,0}+a_{4,1}s+a_{4,2}s^2+a_{4,3}tu\\
A_6^{\rm pol}(s,t,u)&=& 
a_{6,0}+a_{6,1}s+a_{6,2}s^2+a_{6,3}tu+a_{6,4}s^3+a_{6,5}stu
\end{array}\end{equation}
The non-polynomial parts are sums of products of polynomials with analytic 
functions of a single variable $s$, $t$ or $u$ exhibiting the cut 
$[4,\infty)$. At fixed $a_{2,0}$ and $a_{2,1}$, the parameters appearing in 
these non-polynomial terms are linear in the $a_{4,\alpha}$, $\alpha=0,1,2,3$. 
The $2n$-th order term $P_{i,2n}$ of the polynomial $P_i$ defined in 
(\ref{4eq4}) is either a constant or a polynomial of first degree.
\begin{equation}\label{4eq11}
\begin{array}{lll}
P_{0,2}=\alpha_{0,2}, & P_{1,2}=\alpha_{1,2}, & P_{2,2}=0,\\[2mm]
P_{0,4}=\alpha_{0,4}+\beta_{0,4}(x-x_s), & P_{1,4}=\alpha_{1,4}, & 
P_{2,4}=\alpha_{2,4},\\[2mm]
P_{0,6}=\alpha_{0,6}+\beta_{0,6}(x-x_s)+\gamma_{0,6}(y-y_s),\quad & 
P_{1,6}=\alpha_{1,6}+\beta_{1,6}(x-x_s),\quad & 
P_{2,6}=\alpha_{2,6}.
\end{array}\end{equation}
The $\alpha_{i,2n}$, $\beta_{i,2n}$, $\gamma_{i,2n}$ are linear combinations 
of the $a_{2n,m}$.

Now my proposal is to apply the strategy commonly used when dealing with 
perturbation expansions and to require that equation~(\ref{4eq6}) be satisfied 
{\it order by order}, i.e. by 
the leading order term, $n=1$, the sum of this term and the next-to-leading 
order correction $n=2$ and finally the sum of the three first order terms, 
$n=1,2,3$. It might be that assumption(ii) above holds good in an interval 
$(4,\Lambda^2)$ whose maximal size depends on $2n$, the order of the 
approximation. For simplicity I look at the structure of the conditions 
obtained with a fixed $\Lambda^2$.

The polynomial $P_i$ on the left-hand side of (\ref{4eq6}) becomes
\begin{equation}\label{4eq12}
P_i^{(2n)}=\sum_{p=1}^n \lambda^{2p}P_{i,2p},\qquad n=1,2,3
\end{equation}
and the truncation of $H_i$ and $H_i^\chi$ leading to polynomials $Q_i^{(2n)}$ 
and $Q_i^{\chi(2n)}$ has to be adapted to the degree of $P_i^{(2n)}$. The 
$Q_i^{\chi(2n)}$ have the same form as the $P_i^{(2n)}$: they are  
obtained from (\ref{4eq11}) by replacing $\alpha_{i,2n},\dots$ by 
new coefficients $\alpha_{i,2n}^H,\dots$ which are linear in the 
$a_{4,\alpha}$ at fixed $a_{2,0}$ and $a_{2,1}$.

Two $n=1$ conditions are obtained from (\ref{4eq6}):
\begin{equation}\label{4eq13}
\alpha_{i,2}+\alpha_{i,2}^H=G_i(x_0,y_0)+H_i(x_s,y_s),\qquad i=0,1.
\end{equation}

As $x_0$ and $y_0$ are such that $G_i(x_0,y_0)$ cannot be obtained from the 
chiral amplitudes, these conditions do not constrain the chiral parameters in 
a useful way. This confirms the fact that high-energy data alone cannot 
discriminate between the standard~\cite{Bij} and the generalized~\cite{Kne2} versions 
of chiral perturbation theory.

For $n=2$ we have four conditions:
\begin{equation}\label{4eq14}
\begin{array}{rcll}
\alpha_{i,4}+\alpha_{i,4}^H&=&0, &i=0,1,\\[2mm]
\lambda^4\left(\alpha_{2,4}+\alpha_{2,4}^H\right)&=&H_2(x_s,y_s), &\\[2mm]
\lambda^4\left(\beta_{0,4}+\beta_{0,4}^H\right)&=&(\partial H_0/\partial x)(x_s,y_s) &
\end{array}
\end{equation}
and $n=3$ gives six conditions:
\begin{equation}\label{4eq15}
\begin{array}{rcll}
\alpha_{i,6}+\alpha_{i,6}^H&=&0, &i=0,1,2,\\[2mm]
\beta_{0,6}+\beta_{0,6}^H&=&0, &\\[2mm]
\lambda^6\left(\beta_{1,6}+\beta_{1,6}^H\right)&=&(\partial H_1/\partial x)
(x_s,y_s), &\\[2mm]
\lambda^6\left(\gamma_{0,6}+\gamma_{0,6}^H\right)&=&(\partial H_1/\partial y)
(x_s,y_s). &
\end{array}
\end{equation}

The end result is a system of 10 equations which determine the 10 fourth- and 
sixth-order parameters $a_{4,\alpha}$ and $a_{6,\alpha}$ in (\ref{4eq10}) in 
terms of 4 high-energy quantities. These are integrals over absorptive parts 
$A^I(s,\tau(s))$ evaluated at $\tau(s)\approx -(64/27)/8s(s-4)^2)$, 
$s>\Lambda^2$, very close to the forward direction.

Similar constraints have been derived in~\cite{Pen} and~\cite{Kne2}. I obtain 
more conditions than in~\cite{Kne2} because the parameters appearing there are 
not split into second-, fourth- and sixth-order contributions.

The equations~(4.14-15) have not been analyzed in detail to date: this is 
beyond the scope of the present work. This means that I end up with a proposal 
whose practicality remains to be explored.

\appendix
\section{Constructing the polynomial $P_i$: a sample calculation}

The polynomial part of $A^\chi$ produces a polynomial part of $G_i^\chi$ which 
appears unchanged in the polynomial $P_i$ defined in (\ref{4eq4}). The main 
point is to find out the contribution to $P_i$ coming from the non-polynomial 
terms of $A^\chi$. Up to sixth order, these terms have a simple structure, 
some of them having the form
\begin{equation}\label{a1}
\tilde{A}(s,t,u)=R(s)f(s),
\end{equation}
where $R$ is a polynomial and $f$ an analytic function with a right-hand cut 
$[4,\infty)$. As an illustration I compute the terms of $P_0$ and $L_0^\chi$ 
coming from $\tilde{A}$. This produces the following term of $G_0^\chi$:
\begin{equation}\label{a2}
\tilde{G}_0^\chi(s,t,u)={1\over 3}\left(R(s)f(s)+R(t)f(t)+R(u)f(u)\right).
\end{equation}
The functions $f$ of the one- and two-loop amplitudes obey once-subtracted 
dispersion relations. This allows a decomposition of $f$ into a low- and a 
high-energy component:
\begin{eqnarray}
f(s)&=&f_{\rm L}(s)+f_{\rm H}(s), \label{a3}\\[2mm]
f_{\rm L}(s)&=&f(0)+{s\over \pi}\int_4^{\Lambda^2}{{\rm d}\sigma\over 
\sigma}\,{\,\Im f(\sigma)\over \sigma-s}, \label{a4}\\
f_{\rm H}(s)&=&{s\over \pi}\int_{\Lambda^2}^\infty{{\rm d}\sigma\over 
\sigma}\,{\,\Im f(\sigma)\over \sigma-s}. \label{a5}
\end{eqnarray}

The high-energy term $\tilde{H}_0^\chi$ is simply obtained by replacing $f$ by 
$f_H$ in (\ref{a2}). Equation~(\ref{4eq4}) becomes
\begin{equation}\label{a6}
\tilde{P}_0(x,y)={1\over 3}\left(R(s)f_{\rm L}(s)+R(t)f_{\rm L}(t)+
R(u)f_{\rm L}(u)\right)-L_0^\chi(x,y)+G_0(x_0,y_0).
\end{equation}
Inserting the representation~(\ref{a4}) and introducing 
Disc~$\tilde{G}_0^\chi=(1/3)R(s)\,\Im f(s)$ into (\ref{4eq2}) gives an explicit 
expression for the polynomial $\tilde{P}_0$:
\begin{equation}\label{a7}
\tilde{P}_0=C+{1\over 3}f(0)\left[R(s)+R(t)+R(u)\right]+{1\over 
\pi}\int_4^{\Lambda^2}{{\rm d}\sigma\over 
\sigma}\left[S(s,\sigma)+S(t,\sigma)+S(u,\sigma)\right]\Im\!f(\sigma),
\end{equation}
where $S(s,\sigma)$ is a polynomial in two variables:
\begin{equation}\label{a8}
S(s,\sigma)={sR(s)-\sigma R(\sigma)\over s-\sigma}
\end{equation}
and the constant $C$ is given by
\[ C={1\over 3\pi}\int_4^{\Lambda^2}{\rm d}\sigma\left[{1\over \sigma-
s_0}+{1\over \sigma-t_0}+{1\over\sigma-u_0}\right]R(\sigma)\,\Im f(\sigma).\]
For convenience, the Mandelstam variables are used instead of $x$ and $y$.

The contributions of the term $\tilde{A}$ of $A^\chi$, defined in (\ref{a1}), 
to $P_1$, $P_2$, $H_1^\chi$ and $H_1^\chi$ are obtained in a similar way, and 
so are the contributions of the other non-polynomial terms of $A^\chi$.

\section{An upper bound for the Taylor coefficients $C_{n,m}$}

Replace (\ref{3eq9}) by the linear relation
\begin{equation}\label{b1}
{\rho_x\over \rho_1}+{\rho_y\over \rho_2}=1,
\end{equation}
$0\leq\rho_x\leq\rho_1$, where $\rho_1$ and $\rho_2$ are such that 
$D_x(\rho_x)\times D_y(\rho_y)$ belongs to $M_H$ if $\rho_x$ and $\rho_y$ obey 
(\ref{b1}). The Taylor coefficients $C_{n,m}$ defined in (\ref{4eq7}) have the 
upper bound
\begin{equation}\label{b2}
|C_{n,m}|<{\rm Inf}{K\over \rho_x^n\rho_y^m}={K\over 
\rho_1^n\rho_2^m}\left(1+{m\over n}\right)^n\left(1+{n\over m}\right)^m.
\end{equation}
If one chooses $\Lambda^2=16$ and $x_0=-50$, one can take $\rho_1=12.3$ and 
$\rho_2=49.3$. This gives an extremely rapid decrease if $m$ increases, $n$ being 
fixed.

\pagebreak
\noindent{\Large\bf Figure captions}
\begin{description}
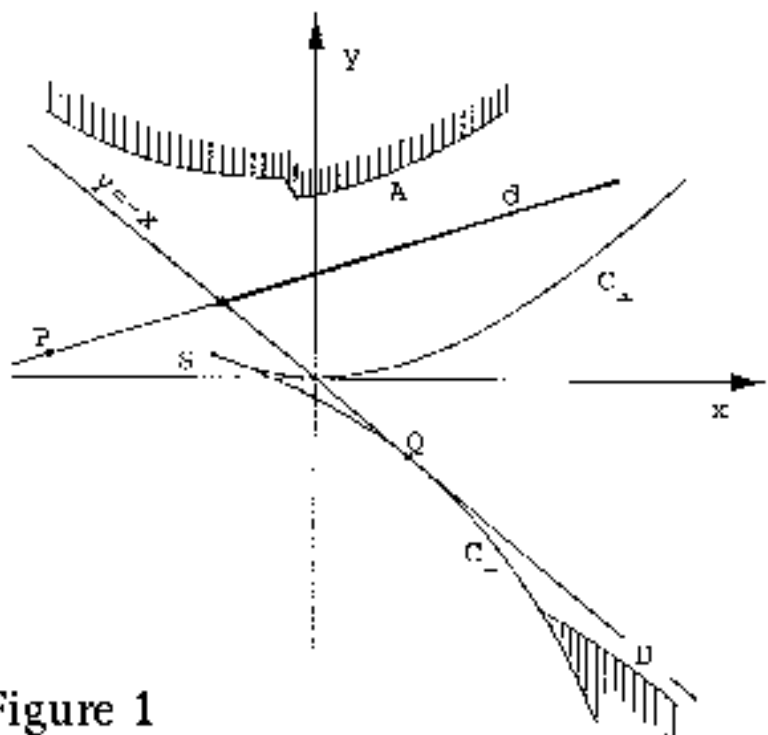
\item[Figure 1:] Qualitative picture of the real $(x,y)$-plane. The real 
$(s,t,u)$-space is mapped onto a domain bounded by the curves $C_+$ and $C_-$ 
($C_-=$~image of the line $s=t$, $t>4/3$, $C_+=$~image of the line $u=t$, 
$t>4/3$). The point $s$ corresponds to the symmetry point $s=t=u=4/3$. The 
line $y=-x$ is the image of $s=4$. The curves $B$ and $A$ are the images of 
the extremities of the semi-major and semi-minor axes of the ellipses $E(s)$ 
($s>4$). For a real slope $a$ and a real point $P(x_0,y_0)$, the restriction 
$F_i$ to the line $d$ ($y=a(x-x_0)+y_0$) has a real cut starting on the line 
$y=-x$ if the point $Q$ is below $d$. It starts on $C_-$ if $Q$ is above $d$. 
The dispersion relation (\ref{2eq6}) is valid if $d$ avoids the shaded region.
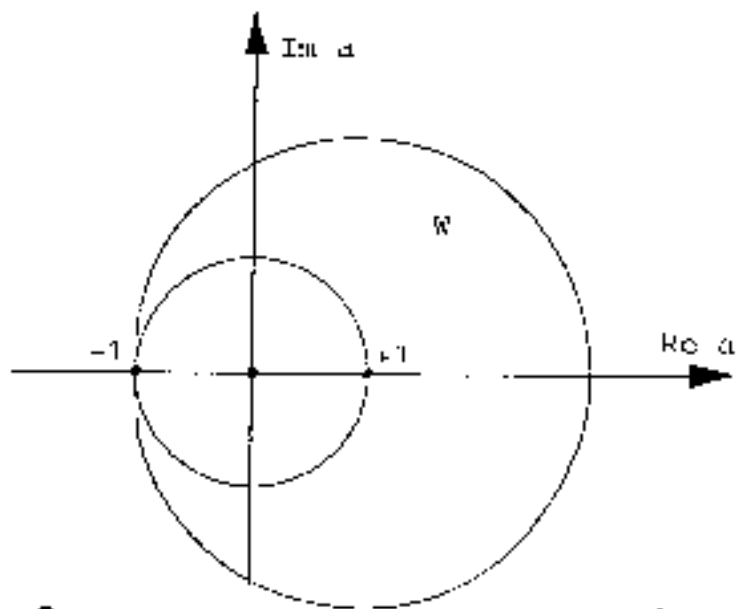
\item[Figure 2:]Domain $W$ of the permitted values of the slope $a$ defined via 
(\ref{2eq14}) for $x_0=-50$ and $y_0=1/27$. The dispersion relation 
(\ref{2eq6}) is valid if $a$ is within this domain; it contains the circle 
$|a|=1$.
\end{description}

\end{document}